\documentstyle[aps,prl,multicol,psfig]{revtex}

\newcommand{\bi}[1]{\bibitem{#1}}
\newcommand{\be}{\begin{eqnarray}}
\newcommand{\ee}{\end{eqnarray}}

\newcommand{\cJ}{ {\cal J} }
\newcommand{\cY}{ {\cal Y} }

\newcommand{\half}{ \frac{1}{2} }

\title{Unification and the Hierarchy from AdS5}

\author{Lisa Randall$^1$\thanks{randall@mit.edu} 
and Matthew D. Schwartz$^{12}$\thanks{matthew@feynman.princeton.edu}
}
\address{$^1$Massachusetts Institute of Technology, Cambridge, MA}
\address{$^2$Joseph Henry Laboratories, Princeton University, Princeton, NJ 08544}

\begin{document}
\maketitle



\begin{abstract}
In AdS$_5$, the coupling for bulk gauge bosons
runs logarithmically, not as a power law. We show
that in the warped scenario addressing the hierarchy
that one can perform calculations, even above
the weak scale, perturbatively.
One can preserve perturbative unification
of couplings. Depending on the cutoff, this can
occur at a high scale. We discuss subtleties
in the calculation and present a regularization
scheme motivated by the holographic correspondence.
We find that generically, as in the standard model,
the couplings almost unify.
For specific choices of the
cutoff and number of scalar multiplets, there is good agreement
between the measured couplings and the assumption of high scale  unification.

\hfill\mbox{[MIT-CTP 3175] }

\end{abstract}


\begin{multicols}{2}[]


\section{Introduction}

The hierarchy problem concerns the discrepancy between the Planck
scale and the electroweak scale, which is roughly sixteen orders
of magnitude. It is difficult to accept this as  a very small
parameter in some fundamental Lagrangian, because quantum
corrections would produce a GUT scale mass for the Higgs and hence
the whole standard model. Solutions to the hierarchy problem
include theories with supersymmetry or technicolor, or, more
recently, extra dimensions. The particular model that we explore
here and in \cite{us2}
involves a single warped extra dimension\cite{rs1}. The
enormous ratio of the two scales arises naturally from an
exponential. There are two apparent weaknesses of theories that use
extra dimensions to produce the hierarchy. The first is that
dangerous operators, such as those that violate baryon number, can
occur with a much larger coefficient than in 4D theories, as the
natural scale is TeV rather than $M_{Pl}$, and has
been addressed in several places \cite{bary1,bary2,bary3}. The second weakness is
that it appears that we must abandon an intriguing feature of
supersymmetric GUT models, that the gauge couplings appear to unify
near the Planck scale.
There have been suggestions for addressing this problem in the
large flat extra dimensional models. For example,
discussed in the context of power law running \cite{dienes1}, 
and exploiting the large extra
dimensions in Ref. \cite{neal}.
Even if these mechanisms were to yield unification,
it would never be at a scale higher than a TeV, because none exists.

In this respect, warped dimensions are very interesting. Although
locally (in the fifth dimension), we see the TeV scale as the
scale at which gravity becomes strongly interacting, this is not
true for the global theory. Scales as high as the Planck
scale appear, but in a different place in the fifth dimension.
Gauge bosons that live in the bulk exist for energies about a TeV,
and can be weakly coupled over the entire range of energies.
Although this might seem surprising for a TeV brane observer, it
is not at all perplexing from the perspective of an observer
on the  Planck brane.

The apparent strong coupling above the TeV scale
appeared to be an obstacle to perturbative
calculations in this regime.
We demonstrate that this is an
artifact of an effective theory calculation, and
one can perform perturbation theory in the full five-dimensional theory.
We regard as a major advance that one can think about physics
above the TeV scale.

We will show that the couplings run with $\beta$-functions
that are essentially a multiple of the standard model
beta functions. This, of course, ensures unification only
with the correct U(1) normalization.  
We will assume the normalization
as in an SU(5) GUT \cite{georgiglashow}. There are many possible ways this
can occur, motivated by a unification group or string theory.
We do not address this issue here in any detail.

In this paper, we show that unification of couplings
is readily achieved in theories with bulk gauge bosons
and warped extra dimensions, with the assumption
of the U(1) normalizaton given above. The details of the precision
with which unification occurs is model dependent, but
generically, the couplings unify at the level of the standard
model, and higher precision is possible with supersymmetry
or additional scalars. The scale at which unification
occurs depends on the cutoff scale where the theory becomes
strongly interacting, which is again a model dependent
parameter, though ultimately one would hope to understand
the microscopic physics sufficiently well to pin it down.
What is clear that even if we view unification as  a clue
to physics underlying the standard model,
there are many possible solutions. The physics of
the warped  models that address the hierarchy
is entirely different from the physics of supersymmetric models.
For example, the particle content at intermediate scales is much richer
than the SUSY desert. The unification of couplings can still be accomodated,
possibly even at a high scale.

Ref. \cite{pom} also considered unified theories with bulk gauge
bosons. However, there the standard model fields were put
on the Planck brane,
not the TeV brane. We abandon this assumption in favor of theories
that directly addres the hierarchy.
Furthermore, we have a different 
regularization and calculational scheme, which we will argue is essential.

\section{Setup and Generalities}
We begin by postulating the presence of a fifth dimension, and an anti-deSitter space metric:
\be
ds_5^{2}=\frac{1}{k^2 z^2}(dt^2 - dx^2 - dz^2) \label{rsmetric}
\ee
The fifth dimension is bounded by two four-dimensional subspaces:
the Planck brane at $z=1/k$ and the TeV brane at $z = 1/T$.
$T$ is related to the size of the extra dimension $R$ by $T = k \exp(-kR)$,
and defines the energy scale on the TeV brane. If we take $T \approx$ TeV,
we can naturally explain the weak scale in the standard model if
the standard model fermions and Higgs are confined to the TeV brane.
Since the fifth dimension is finite, it can be integrated out
to get an effective four-dimensional theory valid at energies below $T$.

Now we put gauge bosons in the bulk \cite{pom,chang,hs,gpom,pombulk,davoudiasl}. 
We can perform a Kaluza-Klein decomposition,
so that the 5D field looks like a tower of 4D fields in the effective theory.
In particular, there is one massless state whose KK profile
is constant in the fifth dimension. Its coupling is the observed 4D coupling,
and the quantity whose running we are interested in calculating. Naively,
one might try running the coupling using a four-dimensional
effective theory with the full spectrum of Kaluza Klein modes.
Such a calculation would give power law running, as with a flat
extra dimension \cite{dienes1}. However, this is incorrect.
Perturbative calculations in the effective four-dimensional
theory cannot be trusted at scales much greater than $T$,
since the theory becomes strongly coupled.
In fact, this is the reason it was
thought perturbative unification was not possible.

However, it is clear that one can do perturbation
theory in the five-dimensional theory up to high scales.
In five-dimensions, where we know that physics
is nonrenormalizable, the theory is ill-defined
at  high energy. We will see this corresponds to the fact
that there is cutoff dependence, and correspondingly regulator
dependence, in our result. But because the background is strongly
curved, there is a large logarithmic
running which is completely calculable. Additional threshold
effects from power law running above the curvature scale
do occur, but they are subleading.

We conclude that we are forced into a full five-dimensional calculation.
From the effective theory or TeV brane perspective, there is a low
cut-off scale on a four-dimensional calculation. On the other hand,
even if one calculates gauge exchange on the Planck brane, KK intermediate
states require a five-dimensional analysis. 
In principle, one can do a four-dimensional holographic
calculation 
\cite{maldacena,witten,gubser,holo,rz,ami,ami2}; however the theory would be strongly coupled
and furthermore one would need a detailed model for the TeV brane.
Below, we present
our method for performing the calculation in the five-dimensional theory. 

\section{Field theory in AdS$_5$}
To study the 5D theory, we will work in position space for the
fifth dimension, but momentum space for the other four. We find it
convenient to do the gauge theory calculations in the Feynman-t'Hooft
gauge. There, the propagator for a gauge boson is:
\be
\langle A^\mu A^\nu\rangle = -i G^{1,0}_p(z,z')\eta^{\mu\nu}
\ee
where the Green's function $G^{1,0}_p$ satisfies:
\be
\left[ \partial_z^2 - \frac{1}{z}\partial_z + p^2\right]
G^{1,0}_p(z,z') =  z k \delta(z-z') \label{eomprop}
\ee
If the gauge boson has positive parity under the orbifold transformation, it must satisfy
Dirichlet boundary conditions at the two branes. Then, the solution is:
\be
G^{1,0}_p(u,v) &=& \left(\frac{\pi k u v}{2}\right)
\frac{ -\cY_0(\frac{p}{k})\cJ_1(p u) + \cJ_0(\frac{p}{k})\cY_1(p u) }
{-\cY_0(\frac{p}{k})\cJ_0(\frac{p}{T})+\cJ_0(\frac{p}{k})\cY_0(\frac{p}{T})} \times\\
&&(-\cY_0(\frac{p}{T}) \cJ_1(p v) + \cJ_0(\frac{p}{T}) \cY_1(p v) )
\ee
where $\cJ$ and $\cY$ are Bessel functions.
The $z$ component of the gauge field, $A_5$,  has a propagator which has the
tensor structure of a scalar, but the Green's function of a vector boson
with bulk mass $M^2 = -k^2$. That is,
\be
\langle A_5 A_5 \rangle = i G^{1,i}_p(z,z')
\ee
where $G^{\sigma,m}_p(z,z')$ is the Green's function for an arbitrary bulk field. In our
notation, $\sigma=2$ for a scalar, $\sigma = 1$ for a vector, and $\sigma=\half$
for spin $\half$. The parameter $m$ is related to the bulk mass $M$ by $M=m k$.

Feynman graphs are to be evaluated by integrating over position in the fifth dimension and
momentum in the other four. Vertices get factors of $(kz)^n$ which come from factors
of the metric in the original Lagrangian. For example, the 4-boson vertex is:
$\sqrt{G} g^{KL} g^{MN} A_K A_L A_M A_N = \frac{1}{kz} A^4$ and so it gets a factor
of $\frac{1}{kz}$, while a $\sqrt{G}\phi^4$ term would have $(kz)^{-5}$. We have to
use the 5D gauge coupling $g_{5D}$ whose square has dimensions of length. It is
related to the 4D coupling by $g_{5D}^2 = R g_{4D}^2$, where $R = k^{-1}\log(k/T)$
is the length of the fifth dimension.

As we discuss in more detail in \cite{us2},
we know the 5D propagator we have
derived cannot be trusted for $qu \gg 1$.
This is to be expected if the physical cutoff
is around $k$, since the effective cutoff will scale with position in the fifth dimension.
In our analysis, we assume the cutoff $\Lambda$ is greater
than $k$ as is expected and necessary for consistency.
In the calculation, the scale $\Lambda$ always appears multiplied
by the warp factor at a given position in the fifth dimension,
so that effectively there is a position-dependent cutoff.
 The obvious way
to implement this cutoff is to integrate up to momentum $q =
\Lambda/(kz)$ at a point $z$ in the bulk. This is almost correct.
As we will argue more fully in \cite{us2}, the correct procedure
is to impose boundary conditions at the scale $\Lambda /(k q)$ on
the Green's functions that appear in the Feynman graphs. That is,
we work with a brane at the Planck scale, and a second brane at
$z=\Lambda /(k q)$. With the second brane at an energy-dependent
position, we integrate out the high energy modes to derive a 5D
Wilsonian effective action valid at energy $q$. Including the warp
factor in the cut-off, as is necessary from general covariance,
guarantees logarithmic running of the coupling as in four
dimensions. Moving the brane in (or equivalently, imposing
$q$-dependent boundary conditions) can be viewed as a choice of
regulator. It is important to recognize that the answer is indeed
cutoff and regulator dependent, since the cutoff is where the
theory goes nonperturbative.

Our choice of regulator is in part motivated by the AdS/CFT correspondence 
\cite{maldacena,witten,gubser,holo,rz}.
The idea is that
string theories in certain AdS backgrounds are probably
dual to conformal field theories in
flat backgrounds. For example, the global symmetry group in AdS$_5$,
$SO(4,2)$, is isomorphic to the conformal group in four-dimensions. In particular, translations
in $z$ in AdS$_5$ correspond to scale transformations in the CFT. So we might
suspect that integrating out a range of scales in the field theory, that is, performing a
a renormalization group flow. might be equivalent to integrating out a length of the fifth
dimension in the 5D theory \cite{herman,bala}.
Since RG-flows transform the whole action, including the
implicit boundary conditions, it makes sense that the normalization of the Green's
functions should depend on scale.  Furthermore, as we discuss more fully in \cite{us2},
this regulator is necessary to get the correct high
energy contribution from light KK modes.

\section{Gauge Boson Self-Energy \label{secbeta}}
\begin{figure}
\begin{center}
\psfig{file=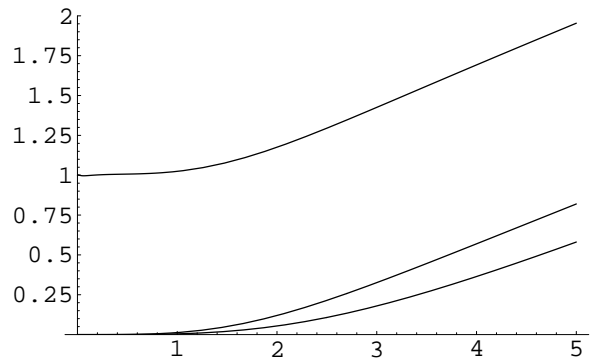,width=3.1in}
\end{center}
\caption{$I_0(\Lambda)$ as a function of $\Lambda/k$.
From top to bottom: massless vector, massive vector (with bulk mass $M=k$),
and Dirichlet vector
(that is, a vector boson with odd parity under the orbifold transformation).}
\label{fig1}
\end{figure}

To simplify the  calculation, we work in the
background field gauge. Here, we summarize
the result and give the details  in \cite{us2}.
The contribution of bulk fields  to the gauge-boson self-energy
is enhanced by a factor:
\be
I^{\sigma,m}(\Lambda,q) = 2 q^4 \int_{1/k}^{\Lambda /(kq)}  \frac{du}{ku}
\int_u^{\Lambda/(kq)} \frac{dv}{kv}
\left[ G^{\sigma,m}_q(u,v)\right]^2
\ee
The functions $I^{\sigma,m}(\Lambda,q)$ are only weakly $q$-dependent, and
can be expanded as $I^{\sigma,m}(\Lambda,q)=I_0^{\sigma,m}(\Lambda)
+I_1^{\sigma,m} {q \over k} + \ldots$.
Numerical results confirm that it is a good approximation
to include only the zeroth
order term $I(\Lambda,q) = I_0(\Lambda)$.
The functions $I_0(\Lambda)^{\sigma,m}$ can be found numerically and are
presented in figure \ref{fig1}.  It is very interesting to examine
this result, as is more fully discussed in \cite{us2}. One
finds the contribution to the beta function from gauge
bosons is essentially multiplied by the number of KK modes
with mass beneath the strong coupling scale. In Ref. \cite{pom},
the curvature scale was above the cutoff, so this was not observed.
If one ignores the fact that there are ghost states and calculates
with Pauli Villars with a higher cutoff, one would find
a similar effect. Notice that although the calculation has
this nice four-dimensional interpretation, it was performed
in the full five-dimensional theory.

Notice also that the coeffficient of the logarithm
scales linearly with $\Lambda$. However,
it is only $\Lambda/k$ that is relevant, not $\Lambda/T$. Moreover,
there is the large logarithm of conventional unification which
is not present for standard power law scaling.
The additional power law corrections that we have omitted
are not logarithmically enhanced, but reflect the nonrenormalizability
of the five-dimensional gauge theory.

Then the 1-loop $\beta$-function is:
\be
\beta(g) =&& -\frac{g^3}{4\pi^2}
\left(C_2(G)(\frac{11}{3}I_0^{1,0}- \frac{1}{6} I_0^{1,i})
 \right. \nonumber\\
&&\quad\quad\quad
-\frac{2}{3} C(r_f) I_0^{\half,0}n_f
\left.-\frac{1}{3} C(r_s) n_s I_0^{2,0}\right)
\ee
The $I_0^{1,i}$ terms come from the contribution of $A_5$, which has the Green's
function of a vector boson with bulk mass  $M^2=-k^2$.
We have included $n_f$ massless bulk Majorana fermions and
$n_s$ massless bulk complex scalars.
$C_2(G)$ is the quadratic group casimir and $C_r$ is the Dynkin index
for the appropriate representations.
Particles which are localized on the TeV brane, such as the matter of the standard model,
only contribute to running below energy $T$ .
However, it is important to keep in mind that any model
with charged fields on the TeV brane requires
bulk charged matter in the bulk. From the holographic viewpoint,
this corresponds to the fact that TeV brane matter are the
bound states of the near CFT theory at higher energy scales.

\section{Coupling Constant Unification \label{secuni}}
There are many possible theories one might consider which include
standard model TeV brane matter and bulk gauge bosons. One essential feature
of any of these models is that baryon number violation
be suppressed. Since the $X$ and $Y$ gauge bosons of an SU(5) model
would necessarily mediate baryon-number violation with a
scale suppression of only a TeV, they must be eliminated.
Two possibilities for eliminating them are either that
they don't exist, or their coupling to standard model
matter is forbidden \cite{cheng,kawa,hall}.
 In either case, there might
be a unified group in a higher dimension or some other
reason to expect a single coupling at high energy. We
simply ask the question, given the low-energy measured
couplings, do they unify at a high energy scale with the assumption
of the U(1) normalization of a GUT model? Of course,
if there is no unified group, for the unification
to be meaningful requires additional physics to
occur at the unification scale.
Otherwise, the lines cross and then diverge at higher
energies.

It should be borne in mind that there is a good deal of
uncertainty in the models. In addition to the question of whether
or not there is a contribution from $X$ and $Y$ gauge bosons,
there is the question of what fermion and scalar fields exist in
the bulk.  We expect there to exist charged fermions and scalars
to explain the fields confined to the TeV brane. We therefore
consider an arbitrary number of scalars and fermions. The scaling
depends relatively weakly on this parameter; depending on the
value, one can obtain very exact unification or unification
roughly at the level of the standard model. It should also be kept
in mind that the threshold corrections to this calculation can be
large since we only focused on the large log term. There are
additional power law corrections between $k$ and $\Lambda/k$, for
example, that can modify our results and should be included in
future work.

For illustration, we  pick a fairly specific but very simple model.
We take $\Lambda=k$ and
put 4 Majorana fermion doublets in the bulk.
These represent the preonic states in the CFT
which condense to form the Standard Model Higgs doublet at low energy.
We should also include bulk fields for the SM fermions.
But since particles which transform in complete $SU(5)$ multiplets do not
affect unification or the unification scale (although they do affect the value of the
couplings at this scale), we simply represent these fields with $n_g=3$ in
the following.. This lets us compare to the SM most easily.
The results for the couplings at a scale $M_G$ are:

\be
\alpha_1^{-1}&&(M_{G})
= \alpha_1^{-1}(M_Z) - \frac{2}{\pi}
\log\left( \frac{M_{G}}{M_Z} \right) \times \nonumber\\
&& \left[\frac{3}{5}\frac{n_f}{12}I_0^{\half,0}(\Lambda) + \frac{n_g}{3} \right]
\nonumber \\
\alpha_2^{-1}&&(M_{G})
= \alpha_2^{-1}(M_Z)
- \frac{2}{\pi}
\log\left( \frac{M_{G}}{M_Z} \right) \times \nonumber \\
&& \left[-\frac{11}{6}I_0^{1,0}(\Lambda) + \frac{1}{12} I_0^{1,i}(\Lambda)
+\frac{n_f}{12}I_0^{\half,0}(\Lambda)+\frac{n_g}{3} \right] \nonumber\\
\alpha_3^{-1}&&(M_{G})
= \alpha_3^{-1}(M_Z) \
- \frac{2}{\pi}
\log\left( \frac{M_{G}}{M_Z} \right) \times  \nonumber \\
&&\left[-\frac{11}{4}I^{1,0}_0(\Lambda) + \frac{1}{8} I_0^{1,i}(\Lambda)
+ \frac{n_g}{3}\right]
\ee

\begin{figure}
\begin{center}
\psfig{file=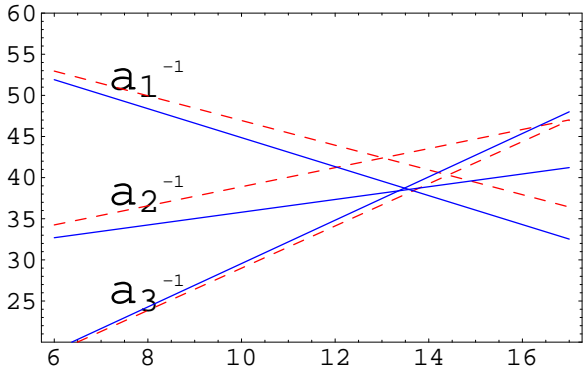,width=3.2in}
\end{center}
\caption{
$\alpha^{-1}$ as a function of $\log_{10} (M_{G}/M_Z)$.
Unification of couplings for $\Lambda = k$ (solid lines). The standard model is shown
for comparison (dashed lines).}
\label{unia5}
\end{figure}

$I_0(\Lambda)$ is roughly equal to the number of KK modes with mass below $\Lambda T$.
It is defined exactly in the previous section. The numerical values are:
$I^{1,0}_0(1) = 1.024$, $I_0^{1,i}(1)=0.147$, and  $I_0^{1/2,0}(1) = 1.009$.
There are additional terms in the above
equations proportional to $I_1(\Lambda) \frac{M_{G}}{k}$, which we will assume to be small.
We will use the observed values \cite{lang} of $\alpha_3(M_Z) = 0.1195$,
$\alpha_e^{-1}(M_Z) = 127.934$, and $\sin^2\theta_w = 0.23107$ at the Z-boson mass $M_Z = 91.187$ GeV.
The couplings are shown for this case in figure \ref{unia5}. The standard model is shown
for comparison.

As $\Lambda$ is increased, $I_0^{1,0}(\Lambda)$ and 
$I_0^{1/2,0}(\Lambda)$ grow at roughly the same rate.
The net effect is that the $\beta$-functions basically scale uniformly with $\Lambda$. This will not have
much of an effect on whether unification occurs, but it can drastically change the scale,
For example, if we take $\Lambda = 5$, the scale drops from $10^{14}$ to $10^8$.
Roughly, the $\beta$-function scales as $(1+(\Lambda-k)/\pi k)$. The
exponent of $(M_G/TeV)$ is divided by this quantity. This
can lead to much lower unification scales as well. Accelerated
unification was considered in a different scenario in Ref. \cite{harvard}.
Their models also have the feature that the unification
scale is connected to observable quantities.

\section{Conclusions}
Unlike in flat extra dimensions, in warped extra dimensions,
parameters run logarithmically. This is because one only sees a
few KK modes; one does not get the sum of all KK contributions
that adds up to power law running \cite{dienes1}. In theories with
a cutoff that is high compared to the curvature scale, this
running is faster than the standard model; if the cutoff is low,
unification is very similar to a four-dimensional model. In this
paper, we have presented a procedure for running couplings in
AdS$_5$, and relating high energy parameters to their value at the
infrared scale (e.g. $\approx$ TeV). This involved a  regularization scheme
motivated by AdS/CFT duality and the Wilsonian effective action.

The original Randall-Sundrum scenario was presented as a solution
to the hierarchy problem. With bulk gauge bosons, it is now clear
that it can also be consistent with coupling constant unification. The
key is that even though gauge bosons are in the bulk, running is
logarithmic, as in  four dimensions. 
Further details and results
are presented in \cite{us2}.

One might argue that the supersymmetry looks better 
from the point of view of unification. 
However, additional threshold corrections are
required even in that case for unification, so the net result is
also model-dependent, especially when one accounts for the absence
of a definite model due to the doublet-triplet splitting problem.
It seems fair to say that both scenarios are possibilities at this
point and that it is premature to deduce knowledge of physics up
to very high energy scales based on unification.

It is clear that there are many possibilities in terms of
models and parameters, and the detailed predictions for the high
energy couplings from the low energy ones will vary. One can for
example consider the supersymmetric version of this theory or
alternative GUT groups. Furthermore, we only include the
logarithmically enhanced contribution. There are further threshold
corrections arising from power law running between $k$ and
$\Lambda$, as well as higher order terms in the expansion of
$I(\Lambda, q)$. These are of course in addition to the standard
subleading corrections. We therefore view this work (and that of
\cite{us2}) as a first step towards a more detailed and more
general analysis.

\section{Acknowledgements}
We thank Neal Weiner for conversations about GUTs that inspired much of this work.
We also wish to thank Nima Arkani-Hamed, Andreas Karch, Emmanuel Katz,
Witek Skiba,
Yasunori Nomura,  Massimo Porrati, and Frank
Wilczek for valuable discussions. We also thank Frank
 for motivation.

\end{multicols}
\end{document}